\documentclass[useAMS,usenatbib]{mn2e}

\usepackage{graphicx}

\def\jh{\mbox{$\rm (J-H)$}}

\def\jk{\mbox{$\rm (J-K_s)$}}
\def\mMJ{\mbox{$\rm (m-M)_J$}}
\def\mv{\mbox{$\rm M_V$}}

\def\ebv{\mbox{$\rm E(B-V)$}}
\def\ejh{\mbox{$\rm E(J-H)$}}
\def\rc{\mbox{$\rm R_{core}$}}

\def\rt{\mbox{$\rm R_{tidal}$}}
\def\rh{\mbox{$\rm R_h$}}

\def\ds{\mbox{$\rm d_\odot$}}
\def\rgc{\mbox{$\rm R_{GC}$}}
\def\feh{\mbox{$\rm [Fe/H]$}}
\def\jj{\mbox{$\rm J$}}
\def\hh{\mbox{$\rm H$}}
\def\ks{\mbox{$\rm K_s$}}
\def\mas{\mbox{$\rm mas\,yr^{-1}$}}

\title[A new Milky Way globular cluster]{FSR\,1767 - a new globular cluster in the Galaxy}

\author[Bonatto et al.]{C. Bonatto$^1$, E. Bica$^1$, S. Ortolani$^2$ and B. Barbuy$^3$\\
$^1$ Departamento de Astronomia, Universidade Federal do Rio Grande do Sul\\
Av. Bento Gon\c{c}alves 9500, Porto Alegre 91501-970, RS, Brazil; charles@if.ufrgs.br;
bica@if.ufrgs.br\\
$^2$ Universit\`a di Padova, Dipartimento di Astronomia\\
Vicolo dell'Osservatorio 5, I-35122 Padova, Italy; sergio.ortolani@unipd.it\\
$^3$ Universidade de S\~ao Paulo, Departamento de Astronomia\\
Rua do Mat\~ao 1226, S\~ao Paulo 05508-900, Brazil; barbuy@astro.iag.usp.br}

\begin{document}


\maketitle


\begin{abstract}
The globular cluster (GC) nature of the recently catalogued candidate FSR\,1767 is established
in the present work. It results as the closest GC so far detected in the Galaxy. The nature of 
this object is investigated by means of 2MASS colour-magnitude diagrams (CMDs), the stellar radial 
density profile (RDP) and proper-motions (PM). The properties are consistent with an intermediate 
metallicity ($\feh\approx-1.2$) GC with a well-defined turnoff (TO), red-giant branch (RGB) and blue 
horizontal-branch (HB). The distance of FSR\,1767 from the Sun is $\ds\approx1.5$\,kpc, and it 
is located at the Galactocentric distance $\rgc\approx5.7$\,kpc. With the space velocity components
$(V,W)=(184\pm14,-43\pm14)\rm\,km\,s^{-1}$, FSR\,1767 appears to be a  Palomar-like GC with 
$\mv\approx-4.7$, that currently lies $\approx57$\,pc below the Galactic plane. The RDP is well 
represented by a King profile with the core and tidal radii $\rc=0.24\pm0.08$\,pc and $\rt=3.1\pm1.0$\,pc, 
respectively, with a small half-light radius $\rh=0.60\pm0.15$\,pc. The optical absorption is moderate 
for an infrared GC, $A_V=6.2\pm0.3$, which together with its central direction and enhanced contamination 
explains why it has so far been overlooked.
\end{abstract}


\section{Introduction}
\label{intro}

With some exceptions, Globular Clusters were formed in the initial phases of the Galaxy and preserve 
information in their structure and spatial distribution that is essential to probe the early Milky Way 
physical conditions. Thus, derivation of the present-day spatial and luminosity distribution of GCs, as 
well as their physical and chemical properties, is important to better understand the formation and 
evolution processes and trace the geometry of the Galaxy (e.g. \citealt{MvdB05}; \citealt{GCProp}).

The number of known Galactic GCs has been slowly increasing as deeper surveys are carried out. The compilation 
of Harris (1996, and the update in 2003\footnote{\em http://physun.physics.mcmaster.ca/Globular.html} - 
hereafter H03) contains 150 members. Later additions to the GC population include the far-IR GC GLIMPSE-C01
(\citealt{KMB2005}), the young halo GC Whiting\,1 (\citealt{Carraro05}), two stellar systems detected with 
the Sloan Digital Sky Survey (SDSS) in the outer halo, SDSS\,J1049$+$5103 (Willman\,1) and SDSS\,J1257$+$3419 
that might be GCs or dwarf galaxies (\citealt{Willman05}; \citealt{Sakamoto06}), and AL\,3, a bulge GC with a 
blue HB (\citealt{OBB06}). Recently, \citet{FMS07} found evidence that FSR\,1735 is a GC in the inner Galaxy, 
and \citet{Belokurov07} found the faint halo GC SEGUE\,1 using SDSS. Finally, \citet{Koposov07} reported 
the discovery of two very-low luminosity halo GCs (Koposov\,1 and 2) detected with SDSS. 
 
In a recent observational effort to uncover potential star clusters, \citet{FSR07} carried out
an automated search for stellar overdensities using the 2MASS\footnote{{\em
http://www.ipac.caltech.edu/2mass/releases/allsky/}} database for $b<20^\circ$, which resulted in a 
list of 1021 candidates. Based on diagnostic diagrams involving number of stars, core radius and central 
density, they classified 9 of these as GC candidates. In the present work we investigate the nature of 
the GC candidate FSR\,1767 by means of 2MASS CMDs, proper motions and a detailed structural analysis.

In Sect.~\ref{PhotPar} we analyze near-IR CMDs, proper motions and cluster structure. In Sect.~\ref{Disc}
we discuss cluster properties. Concluding remarks are given in Sect.~\ref{Conclu}.

\section{Photometric parameters} 
\label{PhotPar}

\begin{figure}
\begin{minipage}[b]{1.0\linewidth}
\includegraphics[width=\textwidth]{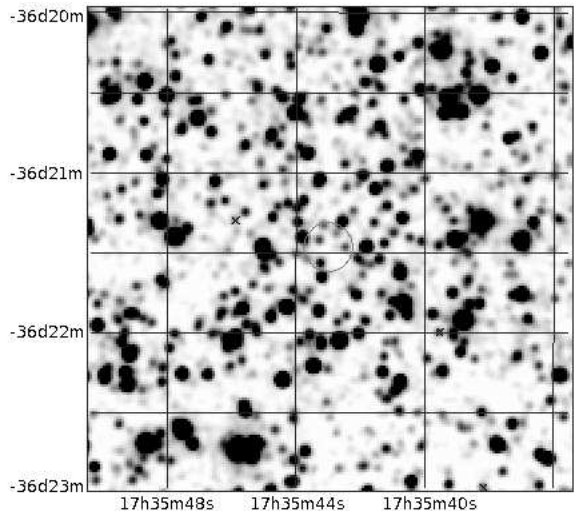}
\end{minipage}\hfill
\caption[]{$3\arcmin\times3\arcmin$ 2MASS \hh\ image of FSR\,1767. The small circle marks
the central region of FSR\,1767 (Sect.~\ref{PhotPar}). North is up and East is left. }
\label{fig1}
\end{figure}

The coordinates of FSR\,1767 (\citealt{FSR07}) are $\alpha(J2000)=17^h\,35^m\,43^s$ and
$\delta(J2000)=-36^\circ\,21\arcmin\,28\arcsec$, which correspond to the Galactic coordinates
$\ell=352.6^\circ$ and $b=-2.17^\circ$. We are dealing with a $\rm4^{th}$ quadrant cluster projected 
against the bulge in Scorpius. The following analysis is based on \jj, \hh\ and \ks\ 2MASS photometry 
taken from VizieR\footnote{\em http://vizier.u-strasbg.fr/viz-bin/VizieR?-source=II/246} and tools as 
described in e.g. \citet{BB07}. As photometric quality constraint, 2MASS extractions were restricted 
to stars with uncertainties in \jj, \hh\ and \ks\ smaller than 0.25\,mag. A typical distribution of 
errors as a function of magnitude can be found in \citet{BB07}. About $75\%$ of the stars have 
errors smaller than 0.06\,mag. The 2MASS H image of FSR\,1767 (Fig.~\ref{fig1}) shows that field 
star contamination is important.

\begin{figure}
\resizebox{\hsize}{!}{\includegraphics{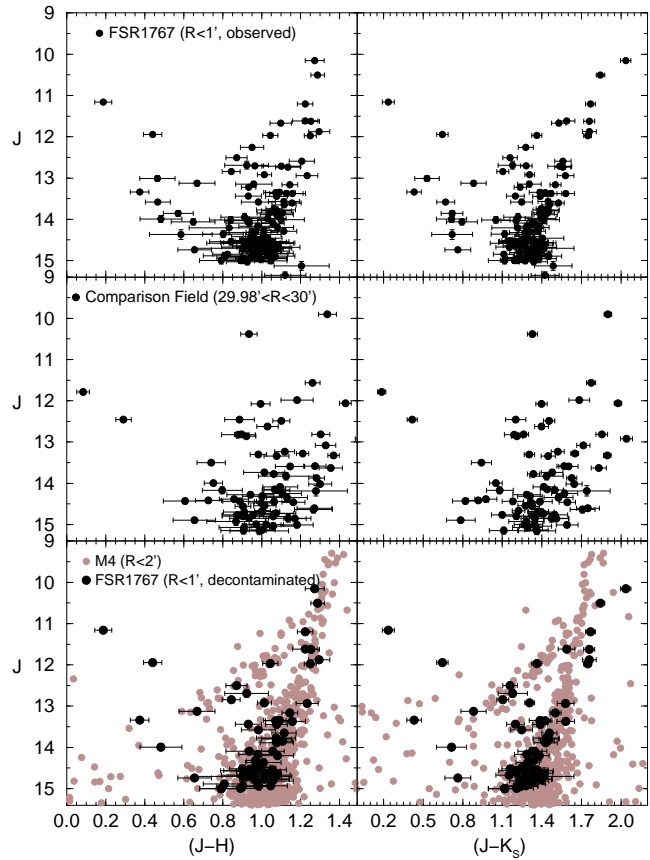}}
\caption{The CMD of the central $R<1\arcmin$\ extraction of FSR\,1767 (top) is compared to an equal 
area offset field extraction (middle). Bottom: the field-star decontaminated $R<1\arcmin$\ CMD 
(filled circles) presents similarities with the $R<2\arcmin$\ CMD of M\,4 (gray). }
\label{fig2}
\end{figure}

Fundamental parameters of FSR\,1767 are derived by comparison with the nearby, intermediate-metallicity
($\ds\approx2.2$\,kpc, $\feh=-1.2$) GC M\,4 (H03, and references therein). To minimize field contamination
effects, we first examine the $\jj\times\jh$ and $\jj\times\jk$ CMDs of a central ($R<1\arcmin$) extraction
around the cluster center (top panels of Fig.~\ref{fig2}). The stellar density profile in this region presents
a high contrast with respect to the background (Sect.~\ref{Struc}). Features displayed by these CMDs are
significantly different from the more scattered ones of the comparison field (middle panels). They resemble
the TO, RGB and HB sequences displayed by the $R<2\arcmin$ region of M\,4 (bottom panels).
The latter CMDs of FSR\,1767 contain statistically significant sequences that 
result from applying the field-star decontamination algorithm (\citealt{BB07}). Basically, the algorithm 
takes into account the relative densities of probable field and cluster stars in small CMD cubic cells with 
sides along \jj, \jh\ and \jk. It is sensitive to local variations of field contamination with colour and 
magnitude. As comparison field we use the region $10\arcmin<R<40\arcmin$ to improve background star-count 
statistics. 

The match between the CMD morphologies of FSR\,1767 and M\,4 results from applying $\Delta\mMJ=0.7\pm0.1$ 
and $\Delta\ejh=0.58\pm0.05$ to M\,4. The match provides for FSR\,1767 a reddening $\ejh=0.63\pm0.03$, which
converts to $\ebv=2.0\pm0.1$, corresponding to $A_V=6.2\pm0.3$\footnote{Reddening transformations are
$A_J/A_V=0.276$, $A_H/A_V=0.176$, $A_{K_S}/A_V=0.118$, and $A_J=2.76\times\ejh$ (\citealt{DSB2002}), with
$R_V=3.1$.}. This absorption coincides with that derived from \citet{Schlegel98} in the direction
of FSR\,1767, which implies that essentially all the reddening arises in the foreground disk. The resulting 
distance from the Sun is $\ds=1.5\pm0.1$\,kpc, which places it as the closest
GC, since among the optical GCs M\,4 and NGC\,6397 are located at $\ds=2.2$ and $\ds=2.3$\,kpc, respectively
(H03). The nearest IR GC is GLIMPSE-C01 at $\ds\approx4$\,kpc (\citealt{KMB2005}). Adopting
$R_\odot=7.2$\,kpc as the Sun's distance to the Galactic center (\citealt{GCProp}), the Galactocentric distance
of FSR\,1767 is $\rgc=5.7\pm0.2$\,kpc, and it is displaced about 57\,pc below the plane. This solution is shown
in the bottom panels of Fig.~\ref{fig2} for both 2MASS colours.

\begin{figure}
\resizebox{\hsize}{!}{\includegraphics{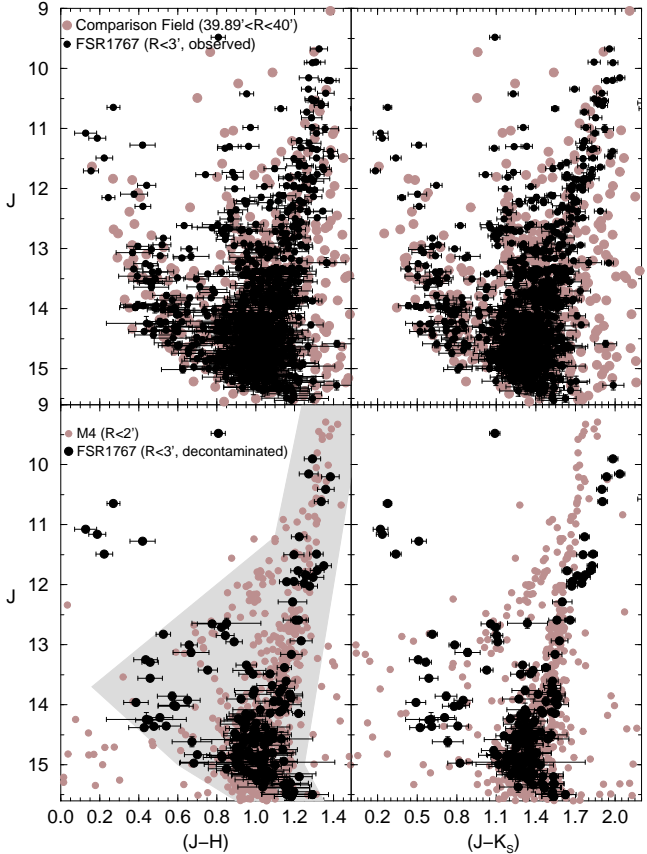}}
\caption{Top: $R<3\arcmin$ CMD of FSR\,1767 (black circles) compared to the equal area CMD of the 
offset field (gray). Bottom: comparison with the $R<2\arcmin$ CMD of M\,4 (gray). Shaded area: 
colour-magnitude filter used to select stars for the radial density profile (Sect.~\ref{Struc}).}
\label{fig3}
\end{figure}

We further analyze the CMD morphology of FSR\,1767 by examining the region $R<3\arcmin$, that basically 
contains the bulk of the cluster stars (Sect.~\ref{Struc}). Likewise in the $R<1\arcmin$ extraction, the 
field decontamination reveals stellar density excesses that correspond to the TO, RGB and HB sequences 
similar to those of M\,4 (Fig.~\ref{fig3}). The results are consistent with an intermediate-metallicity
GC that contains some blue HB stars. In terms of stellar content, FSR\,1767 appears to be similar to the
low-mass GCs AL\,3 (\citealt{OBB06}), Palomar\,13 (\citealt{Siegel01}) and AM\,4 (\citealt{InmanC87}), 
which contain about 10--20 giants. The combined number of RGB and HB stars in M\,4 ($\mv=-7.2$, H03) is 
$\sim10$ larger than that in FSR\,1767. Assuming a similar scaling for the total luminosity and the number 
of giants, we estimate $\mv\approx-4.7$ for FSR\,1767, consistent with its low-luminosity nature.

\subsection{Proper motions}
\label{CentralCMD}

\begin{figure}
\resizebox{\hsize}{!}{\includegraphics{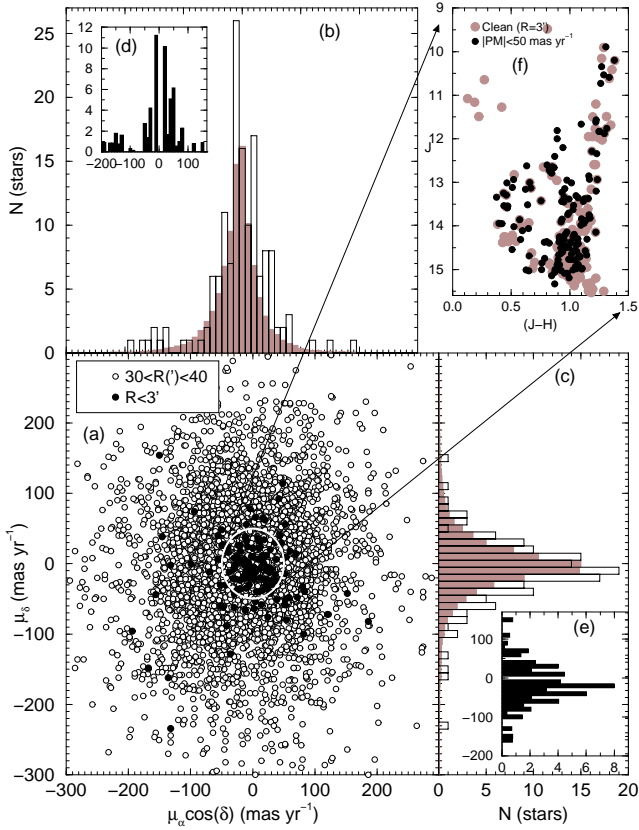}}
\caption{ (a): Proper motion component distribution of the stars within $R<3\arcmin$ (filled 
circles) and those within $30\arcmin<R<40\arcmin$ (empty circles). The white circle isolates low-PM 
stars with $\rm|\mu_\delta|=|\mu_\alpha\cos(\delta)|\la50\,mas\,yr^{-1}$. (b and c): Comparative 
histograms of the $R<3\arcmin$ stars (white) and comparison field (shaded), which was scaled to 
match the projected areas. (d and e): Excess of stars in $R<3\arcmin$ with respect to the comparison 
field. (f): Low-PM stars (filled circles) occur mostly in the MS/TO, RGB and HB sequences of the 
decontaminated photometry (gray circles).}
\label{fig4}
\end{figure}

Another clue to the GC nature of FSR\,1767 is provided by 
NOMAD\footnote{\em http://vizier.u-strasbg.fr/viz-bin/VizieR?-source=I/297} proper motion data 
(Fig.~\ref{fig4}) taken for stars in the same region as the 2MASS data. NOMAD is based on
the International Celestial Reference System (ICRS) whose origin is located at the barycenter of 
the solar system. However, we note that the correspondence between NOMAD and 2MASS is not complete. 
For instance, among the 750 stars detected with 2MASS for $R<3\arcmin$, 173 ($\approx23\%$) are 
included in NOMAD. Most of the $R<3\arcmin$ NOMAD stars form a compact clump, defined by 
$\rm|\mu_\delta|=|\mu_\alpha\cos(\delta)|\la50\,mas\,yr^{-1}$, in the PM components plane (panel a). 
About 75\% of these stars have PM errors smaller than $\rm10\,mas\,yr^{-1}$. Systematic 
differences in the PM component distributions for $R<3\arcmin$ with respect to the comparison 
field are apparent in the respective, nearly Gaussian histograms (panels b and c). Besides different 
values of the distribution maximum in both PM components, there are significant excesses of stars in 
$R<3\arcmin$ with respect to the comparison field at several bins (panels c and d). From Gaussian fits 
to the distributions we measure the average values $\langle\mu_\alpha\cos(\delta)\rangle=2.81\pm2.85\,\mas$
and $\langle\mu_\delta\rangle=-8.78\pm2.82\,\mas$ for stars in the region $R<3\arcmin$, and
$\langle\mu_\alpha\cos(\delta)\rangle=-2.53\pm2.01\,\mas$ and $\langle\mu_\delta\rangle=-3.66\pm2.04\,\mas$ 
for the comparison field. A more physical interpretation on the motion of FSR\,1767 is provided
by the UVW space velocity components, where U is positive towards the Galactic center, V in the direction 
of the Galactic rotation and W towards the North Galactic Pole. We take $U=0$ since radial velocity is not 
available. After correcting for a peculiar Solar motion of
$(U_\odot,V_\odot,W_\odot)=(+10.00\pm0.36,+5.25\pm0.62,+7.17\pm0.38)\rm\,km\,s^{-1}$ (\citealt{DehBin98}),
and taking into account the circular velocity of the Sun, $220\rm\,km\,s^{-1}$, we derive the cluster's space 
motion components $(V,W)=(184\pm14,-43\pm14)\rm\,km\,s^{-1}$ with respect to a stationary Galactic reference 
frame. These velocities were computed using the cluster distance, $\ds=1.5$\,kpc. We conclude that most of
the space motion of FSR\,1767 is prograde and similar to the expected circular velocity at its Galactocentric
distance, suggesting that this could be a GC sharing a disk kinematics. Currently, FSR\,1767 lies below the 
plane moving away 
from it at an angle $\theta=13^\circ\pm4^\circ$. When transposed to the $R<3\arcmin$ CMD (panel f), the PM 
clump stars consistently should belong mostly in the MS/TO, RGB and HB sequences defined by the decontaminated 
photometry. As expected for a star cluster, the MS/TO, RGB and HB stars share a similar kinematics.

\subsection{The structure of FSR\,1767}
\label{Struc}

\begin{figure}
\resizebox{\hsize}{!}{\includegraphics{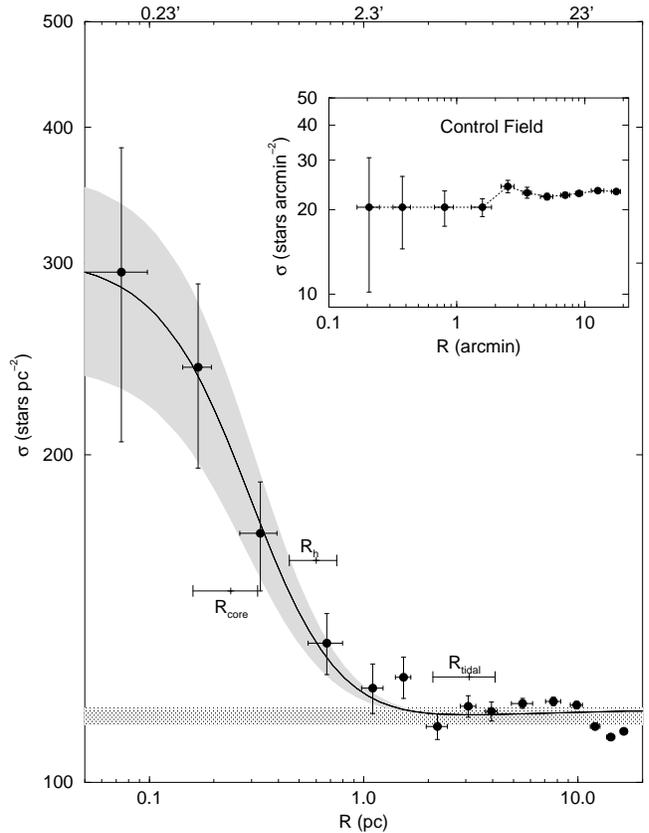}}
\caption{Stellar RDP (filled circles) of FSR\,1767 in absolute scale. Solid line: best-fit King
profile. Horizontal shaded region: offset field stellar background level. Core, half-light and tidal 
radii are indicated. Gray regions: $1\sigma$ King fit uncertainty. The angular scale is in the upper 
abscissa. Inset: RDP of a nearby control field.}
\label{fig5}
\end{figure}

Structural parameters are derived by fitting the \citet{King1966} law to the stellar radial density profile,
built with colour-magnitude filtered photometry for $R<40\arcmin$ (e.g. \citealt{BB07}). The filter
(Fig.~\ref{fig3}) removes contamination of stars with colours deviant from cluster sequences in the CMD. 
The resulting profile (Fig.~\ref{fig5}), shown in absolute units corresponding to $\ds=1.5$\,kpc, presents 
a prominent density excess over the background, especially for $R<2$\,pc ($\approx4.5\arcmin$). Within
uncertainties, most of the RDP is well represented by a King law characterized by a core radius
$\rc=0.24\pm0.08$\,pc ($\approx0.54\arcmin$), and a tidal radius $\rt=3.1\pm1.0$\,pc ($\approx7\arcmin$),
with a concentration parameter $c=\log(\rt/\rc)=1.1\pm0.2$. The present value of \rc\ results the same as 
that in \citet{FSR07}, but their tidal radius is $\approx50\%$ larger than the present one. FSR\,1767 
presents a small half-light radius, $\rh=0.60\pm0.15$\,pc ($\approx1.4\arcmin$), measured in the \jj\ 
band. These values put FSR\,1767 in the small-radii tail of the Galactic GCs distribution (e.g.
\citealt{MvdB05}).

\subsection{Testing a control field}
\label{ControlF}

\begin{figure}
\resizebox{\hsize}{!}{\includegraphics{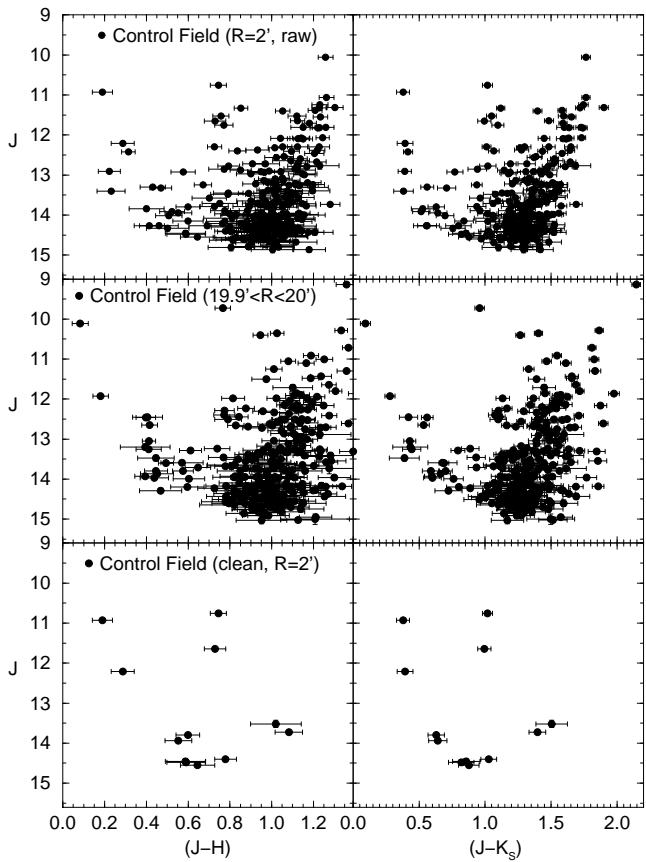}}
\caption{Top: central extraction of a control field taken from a bulge direction. Middle:
equal area extraction taken from the border. Stars that remain in the decontaminated CMDs
(bottom) correspond to the expected density fluctuation of the dense field.}
\label{fig6}
\end{figure}

FSR\,1767 is projected against a dense bulge stellar field that is expected to produce CMDs of an 
old population. This can be clearly seen in the CMDs of a randomly selected control field located 
at $\ell=353.6^\circ$ and $b=-2.17^\circ$, about $1^\circ$ to the east of FSR\,1767 (Fig.~\ref{fig6}).
These CMDs contain basically the bulge TO, giant branch and some disk stars. When applied to the 
central $R<2\arcmin$ extraction, the decontamination algorithm removes most of the stars in the CMDs, 
leaving only a few that reflect the expected statistical fluctuation of the dense field.

To test the hypothesis that the high stellar densities associated with central directions might produce
RDPs that mimic those of star clusters we apply the same algorithm (Sect.~\ref{Struc}) to the control 
field (Fig.~\ref{fig6}). The resulting RDP is characterized by fluctuations typical of a dense stellar
field distribution (inset of Fig.~\ref{fig5}).

\section{Discussion}
\label{Disc}

\begin{figure}
\resizebox{\hsize}{!}{\includegraphics{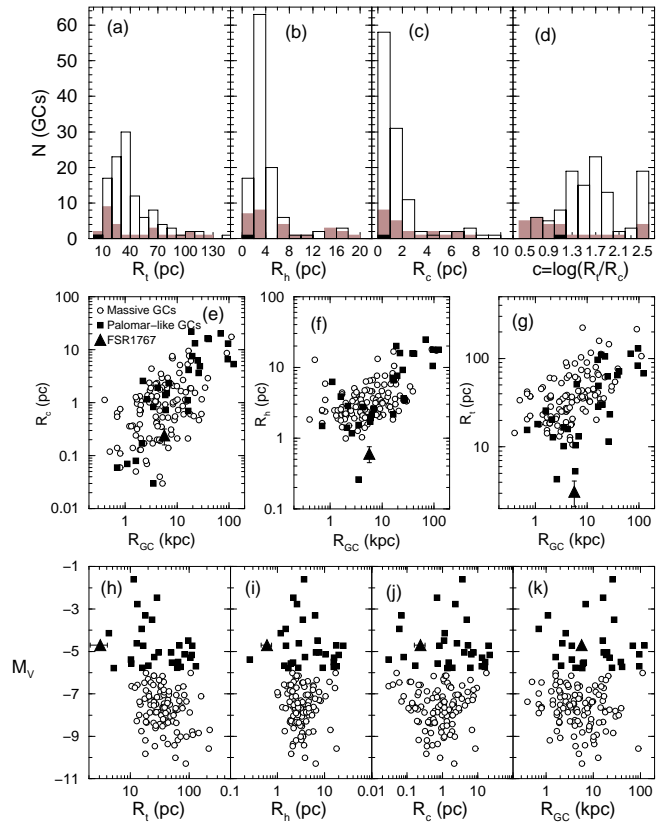}}
\caption{Top: structural parameters of the Palomar-like ($M_V>-6$) GCs (gray-shaded histogram) compared 
to the massive ($M_V<-6$) ones (white). FSR\,1767 is indicated by a black square. Middle: cluster
radii as a function of Galactocentric distance. Bottom: \mv\ as a function of the cluster radii and
Galactocentric distance. FSR\,1767 presents structural properties of a Palomar-like GC. Except for
FSR\,1767, the data are from H03.}
\label{fig7}
\end{figure}

The previous analyses strongly indicate that FSR\,1767 is a Palomar-like GC, a class of GCs that can 
be basically characterized by low mass and $M_V>-6$. H03 contains 30 GCs that satisfy these criteria,
$\approx2/3$ of which are original Palomar GCs. Structural properties of the Palomar and massive
(i.e. non-Palomar) GCs are investigated in Fig.~\ref{fig7} with data from H03. The main difference 
between both classes of GCs is the more concentrated distribution of tidal radii towards smaller 
values occurring for the Palomar GCs (panel a). The distributions of half-light and core radii, on 
the other hand, are similar (panels b and c). Reflecting this, the concentration parameters of the
Palomar-like GCs tend to be smaller than those of the massive ones (panel d). The Palomar-like GCs 
follow the well-known (e.g. \citealt{MvdB05}; \citealt{DjMey94}) relation of increasing cluster radii 
with Galactocentric distance (panels e to g). However, as expected from panel (a), the Palomar-like 
GCs tend to have smaller tidal radii than the massive ones for a given Galactocentric distance, 
especially for small \rgc\ (panel g). The bottom panels show how \mv\ relates with the cluster
radii and Galactocentric distance. In all cases, the structural parameters of FSR\,1767 are 
consistent with those of a low-mass Palomar-like GC dwelling inside the Solar circle, at 
$\rgc\approx5.7$\,kpc.

The small size of FSR\,1767, especially the tidal radius, may be a consequence of its low-mass
nature associated with the inner-Galaxy location, a region with enhanced tidal stress. In the long 
term, dynamical heating of a GC is expected as a consequence of tidal interactions by shocks due to 
disk and bulge crossings as well as encounters with massive molecular clouds. This enhances the rate
of low-mass star evaporation and accelerates the process of core collapse, especially for low-mass
clusters (e.g. \citealt{DjMey94}). One consequence of the evolution of GCs inside the Solar circle 
may be a depopulation of the low-mass tail of the GC distribution.

\section{Summary and conclusions}
\label{Conclu}

FSR\,1767 was first identified as a stellar overdensity projected against the bulge by \citet{FSR07} 
in an automated star cluster survey using 2MASS. They classified it as a GC candidate. Combining 
near-IR photometry and proper motions we conclude that FSR\,1767 is a new GC in the Galaxy. The 
census (\citealt{GCProp}) and recent discoveries (Sect.~\ref{intro}) indicate that FSR\,1767 is 
the $\rm158^{th}$ GC detected in the Galaxy. FSR\,1767 is remarkably close to the Sun at 
$\ds=1.5\pm0.2$\,kpc ($\approx1.5$\,kpc inside the Solar circle), resulting as the nearest GC so far 
detected. It lies close to the Galactic plane at $Z\approx-57$\,pc. According to the proper motion 
properties (Sect.~\ref{CentralCMD}), the space velocity components are 
$(V,W)=(184\pm14,-43\pm14)\rm\,km\,s^{-1}$. Currently, FSR\,1767 is moving 
away from the plane at an angle $\theta=13^\circ\pm4^\circ$. The photometric and structural properties are 
consistent with a Palomar-like GC in the inner Galaxy with $\mv\approx-4.7$, i.e. a low-mass GC containing few 
giants. Its structure is characterized by core, half-light and tidal radii of $0.24\pm0.08$\,pc, $0.60\pm0.15$\,pc 
and $3.1\pm1.0$\,pc, respectively. The TO, RGB and HB of FSR\,1767 are readily detected in the present near-IR 
CMDs (consistent with the proximity), as well as the upper MS and blue HB stars. Based on the similarity with 
the CMD morphology of M\,4, a metallicity $\feh\approx-1.2$ and an absorption $A_V=6.3\pm0.3$ are estimated. 
The Palomar-like nature, projection against the bulge, and the fact that the bulk of its stars require IR 
photometry owing to a relatively high absorption, can explain why FSR\,1767 has been so far overlooked. The 
fact that in less than three years five new Palomar-like GCs have been identified, AL\,3 in the bulge, 
FSR\,1767 at $\rgc\approx5.7$\,kpc, and the three SDSS low-luminosity ones in the halo (Sect.~\ref{intro}), 
suggests that the number of low-mass GCs may be considerably larger than previously thought.

\section*{acknowledgments}
We thank the anonymous referee for interesting suggestions.
We acknowledge partial support from CNPq and FAPESP (Brazil), and MURST (Italy).


\end{document}